\documentclass[12pt]{article}
\usepackage{graphicx}
\title{Static Three-Quark Potential in SU(3) Lattice QCD Monte-Carlo Simulation}

\author{
{T. T. Takahashi, H. Matsufuru, Y. Nemoto and H. Suganuma}
\thanks
{RCNP, Osaka University, Mihogaoka 10-1, Ibaraki 567-0047, Japan.}}

\begin{document}
\maketitle

\section*{Abstract}

The static three-quark(3Q) potential is investigated in the SU(3)
lattice QCD at the quenched level, using the standard Wilson action with 
$\beta =5.7 (a\simeq 0.2{\rm fm})$ and $12^3\times 24$.
With the 3Q Wilson loop, the ground-state 3Q potential $V_{\rm 3Q}$
is calculated for 16 patterns of the 3Q configuration, using the smearing 
technique, which enhances the ground-state overlap of the 3Q operator.
With the accuracy better than a few \%, the lattice QCD data for $V_{\rm 3Q}$
are well described by a sum of the Coulomb term, a constant and the linear
term which is proportional to the minimal length $L_{\rm min}$ of the 
flux-tube linking the 3 quarks.
Our results seem to support the Y-type flux picture rather than $\Delta$-type
one.
From the comparison of $V_{\rm 3Q}$ with the ${\rm Q}$-${\rm \bar{Q}}$
potential, we find the universality of the string tension, 
$\sigma_{\rm 3Q}\simeq \sigma_{\rm Q\bar{Q}}$ as well as the OGE
relation on the Coulomb coefficient, $A_{\rm 3Q}\simeq 
\frac{1}{2}A_{\rm Q\bar{Q}}$.

\section{Theoretical Consideration}

In this paper, we study the 3Q potential $V_{\rm 3Q}$ in the SU(3) lattice 
QCD.
In the inter-quark potential, the short-distance behavior is described with the
one-gluon-exchange (OGE) Coulomb potential originating from P-QCD, 
and the long-distance behavior seems characterized by the flux-tube 
picture, which is derived from the strong-coupling QCD.
For instance, the quark-antiquark (${\rm Q\bar{Q}}$) static potential is 
well described as 
\begin{eqnarray}
V_{\rm Q\bar{Q}}=-\frac{A_{\rm Q\bar{Q}}}{r}+\sigma_{\rm Q\bar{Q}}r
+C_{\rm Q\bar{Q}}.\label{eq1}
\end{eqnarray}
In SU(3) QCD, there appears the `junction' combining three flux tubes
in the baryonic 3Q system, so that the Y-type flux formation is expected.
Then, the 3Q potential $V_{\rm 3Q}$ is conjectured to be described as
\begin{eqnarray}
V_{\rm 3Q}=-\sum_{i<j}\frac{A_{\rm 3Q}}{|\vec{r_i}-\vec{r_j}|}
+\sigma_{\rm 3Q}L_{\rm min}+C_{\rm 3Q},\label{eq3}
\end{eqnarray}
where $L_{\rm min}$ denotes the minimal length of the flux tube linking the 3
quarks.

\section{Numerical Results}

We investigate the 3Q potential $V_{\rm 3Q}$ using the 3Q Wilson loop 
in the SU(3) lattice QCD simulation.
Here, the standard Wilson action is used with $\beta =5.7$ 
$(a\simeq 0.2$ fm) on $12^3\times 24$ lattice at the quenched level.
To enhance the ground-state overlap of the 3Q operator, 
the smearing technique is applied on the spatial link variables.
After the smearing, more than 80\% ground-state overlap is achieved.

We show in Fig.1 the numerical result on $V_{\rm 3Q}$ for 16 patterns 
of the 3Q configuration, where 3 quarks are located at 
$(i, 0, 0), (0, j, 0), (0, 0, k)$ in ${\bf R}^3$.
With the accuracy better than a few \%, the 3Q potential is well reproduced 
by Eq.(\ref{eq3}) with the parameters in Table 1.
(As a comparison,
we try to fit $V_{\rm 3Q}$ with $-\sum_{i<j}\frac{\widetilde{A}_{\rm 3Q}}{|\vec{r_i}-\vec{r_j}|}
+\widetilde{\sigma}_{\rm 3Q}L_{\Delta}+\widetilde{C}_{\rm 3Q}$ with the perimeter $L_{\Delta}$.
In this $\Delta$-type fit, $\chi^2/N_{\rm DF}$ is more than 10.9, 
which seems too large to be acceptable.)

\vspace{1pc}
\hspace{-3pc}
\begin{minipage}{21pc}
\includegraphics[width=1.0\textwidth]{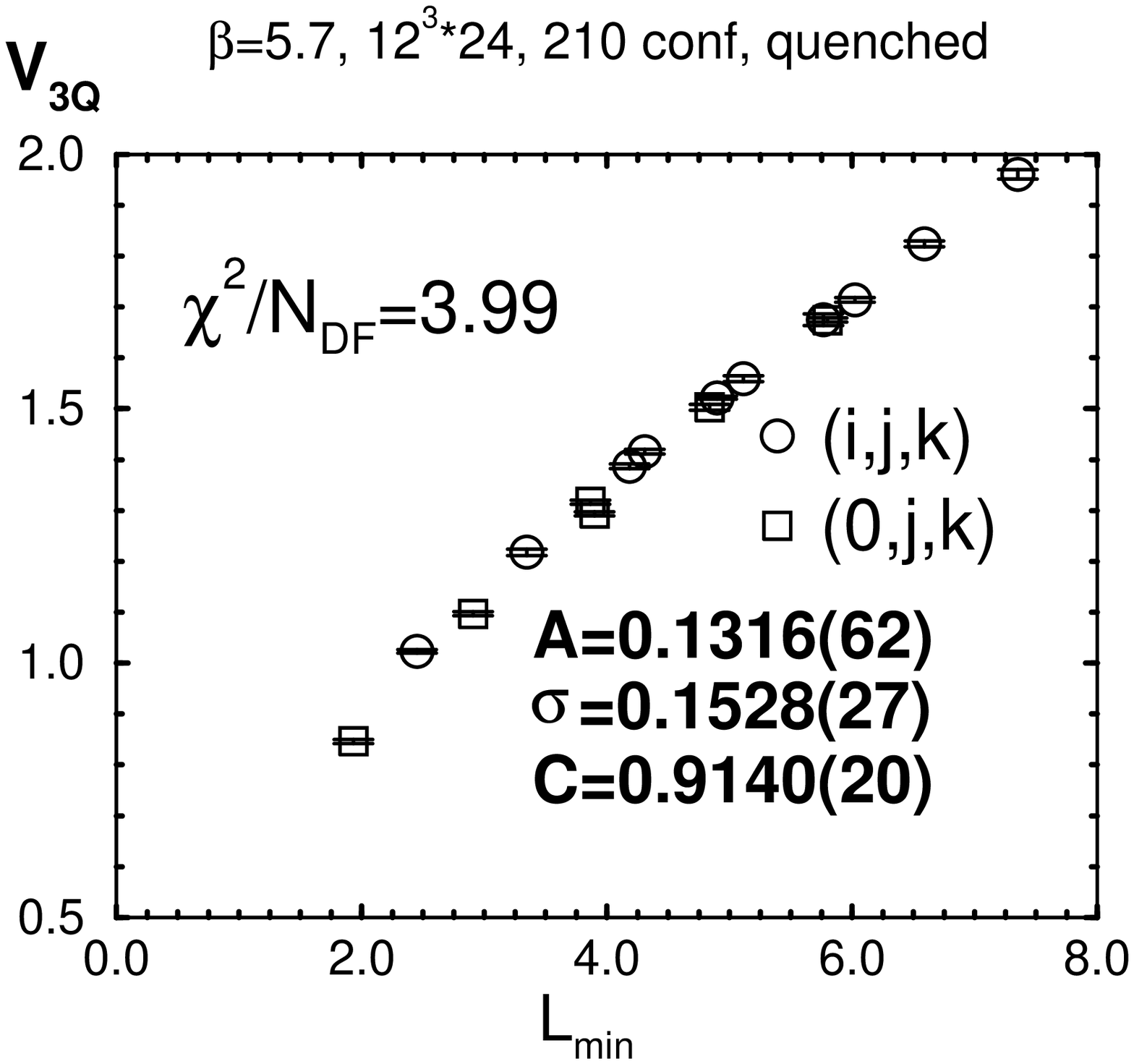}
\end{minipage}
\begin{minipage}{12.5pc}
Fig.1 : \\
3 quark potential $V_{\rm 3Q}$ as the function of $L_{\rm min}$, 
the minimal length of flux linking 3 quarks, in the lattice unit.
Apart from a constant, $V_{\rm 3Q}$ is almost proportional to $L_{\rm min}$
in the infrared region.
\end{minipage}

\vspace{3mm}
\hspace{-3pc}
\begin{minipage}{21pc}
\begin{tabular}{|c||c|c|c|} \hline
 & $A$ & $\sigma$ & $C$ \\ \hline
${\rm Q\bar{Q}}$ & 0.2793(116) & 0.1629(47) & 0.6293(161) \\ \hline
${\rm 3Q}$ & 0.1316(62) & 0.1528(27) & 0.9140(201) \\ \hline
\end{tabular}
\end{minipage}
\begin{minipage}{12.5pc}
Table 1:\\
The Coulomb coefficient, the string tension and the constant term for
${\rm Q\bar{Q}}$ and 3Q potentials in the lattice unit.
\end{minipage}

\vspace{5mm}
Next, we compare the ${\rm 3Q}$ potential with the ${\rm Q\bar{Q}}$ potential
in terms of the fit parameters
($A_{\rm 3Q}, \sigma_{\rm 3Q}, C_{\rm 3Q}$) and 
($A_{\rm Q\bar{Q}}, \sigma_{\rm Q\bar{Q}}, C_{\rm Q\bar{Q}}$) as shown in Table 1.
\begin{itemize}
\item We find the OGE relation on the Coulomb coefficient as 
$A_{\rm 3Q}\simeq \frac{1}{2}A_{\rm Q\bar{Q}}$.
\item The universality of the string tension is found as 
$\sigma_{\rm 3Q}\simeq \sigma_{\rm Q\bar{Q}}$.
\end{itemize}
To conclude, the lattice QCD data for $V_{\rm 3Q}$ are well reproduced 
by Eq.(\ref{eq3}).

\end{document}